# Study of Full Parallel RS(31,27) Encoder for a 3.2 Gbps Serial Transmitter in 0.18 um CMOS Technology

Guangyu Zhang, Quan Sun, Tiankuan Liu, Datao Gong, Dongxu Yang, Yi Feng, Jian Wang, *Senior Member, IEEE*

*Abstract*—This work presents the design of an RS(31,27) Reed Solomon encoder for a 3.2 Gbps serial transmitter in 0.18 um CMOS technology. The proposed encoder is designed with a novel full parallel structure optimized for high speed and high stability. One data frame contains 2 interleaved RS(31,27) codes and thus it can correct at most 20 bits of consecutive errors. A corresponding decoder is implemented on Xilinx Kintex-7 FPGA.

## I. Introduction

Reed Solomon is one of the most powerful forward error correction (FEC) code[1]. It can improve the transmission accuracy with low overhead. It is even suitable for burst error correction, which is common in optical fiber transmission in radiation environment. The RS(31,27) code has 27 information symbols and 4 redundancy symbols. Each symbol contains 5 bits. There are totally 155 bits in one code, including 135 information bits and 20 redundancy bits. One data frame is constructed with 10-bit header and two interleaved RS(31,27) codes, so this codec can correct at most 20 bits of consecutive errors. Each frame contains 320 bits. One must be generated in one 10 MHz clock cycle to reach the speed of 3.2 GHz.

The transmitter is designed for data transmission of a monolithic active pixel sensor (MAPS) in the ATLAS experiment. Triple modular redundancy (TMR) has been implemented to protect the logic from single event upset (SEU). The sensor data is sent from the transmitter ASIC through optical fiber and goes to the decoder running on a Xilinx Kintex-7 FPGA board.

## II. Encoder design

We first tried the regular serial structure and a novel parallel structure to develop the RS encoder. The serial encoder structure in Figure 1 is a LFSR based encoder structure, which is common in many existing designs[2]. The input data port width is 5 bits. This encoder structure needs 31 clock cycles to encode one set of data. The driving clock of the encoder is set to 320 MHz to meet the speed requirement. All of the additions and multiplications are performed in finite field GF(32)[3]. The encoder has to complete a multiplex, a GF(32) multiplication and a GF(32) addition in one cycle (3.125 ns). The circuit cannot meet the timing constraints in our 0.18 um CMOS technology. A way to solve the timing issue is to add another pair of encoders to the transmitter to slow down the clock speed requirement, and a data buffer and an output multiplex are also needed for data routing. This will complicate the design.

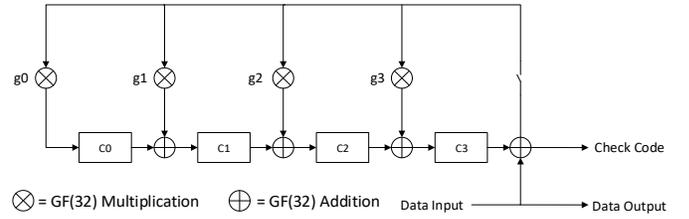

Figure 1 Diagram of LFSR based encoder structure

Instead of adding another pair of encoders, we developed a novel parallel encoder structure. The relationship between information bits and validation bits can be calculated using symbolic computation. A python script uses manual designed symbolic computation to simulate the verilog encoder circuit, runs it for 31 cycles and outputs the generated verilog mapping circuit. The generated circuit is completely a combinatorial logic circuit composed with only XOR operations. The encoder needs only one cycle to encode a data with the clock speed of 10 MHz. Each validation bit is a XOR chain of information bits. The longest XOR chain contains XOR operations for 70 information registers. The XOR chains are synthesized as a tree of XOR3 gates, so the time complexity is O($\log_3 N$), and the deepest tree has a depth of 4. The parallel encoder can be speed up by just increasing clock frequency with no extra resource increment.

Table 1 Resource usage of the two encoder structures

This work was supported by the National Natural Science Funds of China under Grant No: 11773026, 11728509, the Fundamental Research Funds for the Central Universities (WK2360000003, WK2030040064), the Research Funds of the State Key Laboratory of Particle Detection and Electronics.

The authors Guangyu Zhang, Dongxu Yang, Yi Feng, Jian wang are with the University of Science and Technology of China, Jian Wang, State Key Laboratory of Technologies of Particle Detection and Electronics, University of Science and Technology of China, Hefei, Anhui 230026, China (e-mail: Jian Wang, wangjian@ustc.edu.cn).

The authors Quan Sun, Tiankuan Liu, Datao Gong are affiliated with Southern Methodist University, Department of Physics, Southern Methodist University, Dallas, TX 75275, USA.

|  | Parallel Encoder | Serial Encoder |
|---|---|---|
| Total Cells | 1005 | 203 |
| Combinational Cells | 850 | 171 |
| Sequential Cells | 155 | 32 |

The parallel structure costs about 5 times resources than the serial structure. The serial encoder needs to be duplicated to reduce clock frequency, which increases complexity and delay. In consideration of the duplication and the multiplex of the serial encoder, the resource quantity of the parallel structure is not much more than the serial encoder. The serial encoder also needs two data width converters, from 135 bits to 5 bits and from 5 bits to 155 bits. The parallel structure is a good approach for those who use a short Reed Solomon code and want a very fast speed.

### III. SYSTEM DESIGN

A receiver with an RS(31,27) decoder is developed on a Xilinx Kintex-7 FPGA board using the Berlekamp-Massey[4] algorithm. There are totally 4 decoder modules to process the input data. Each two modules form a group. A multiplexer switches the two groups. The clock frequency of a decoder module is 240MHz. A total of 40 clock cycles are needed to decode a code. The transmitter and the decoder board have been integrated tested.

The sensor board connects to the Xilinx Kintex-7 FPGA receiver with optic fiber. The transmitter first scrambles sensor data to achieve DC-balance, and then encodes the data with two RS(31,27) code encoder in parallel. The two data streams are interleaved together to form a 310-bit output port in 10 MHz clock. A frame builder module adds a 10-bit header to the data and converts it to a 32-bit port in 100 MHz clock. Finally, a serializer serializes the 32-bit data stream to a 1-bit double data rate data stream in 1.6 GHz clock. See Figure 2.

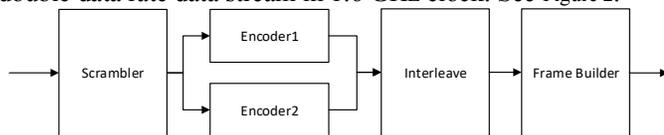

Figure 2 Diagram of Transmitter structure

### IV. TEST RESULT

This design has been tested by tuning the optical modulation amplitude (OMA) using an optical attenuator in the data link. The original code is error-free when the OMA is above -16.9 dBm, while the decoded data is error-free when the OMA is above -16.8 dBm. The Reed Solomon codec successfully recovers 90 incorrect frames at -16.9 dBm.[5]

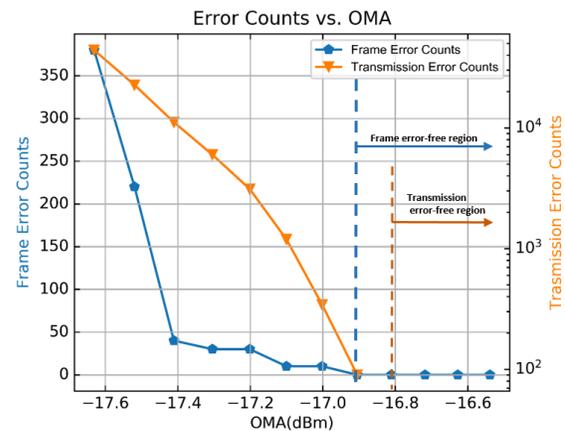

Figure 3 RS(31,27) performance measured with the error counts before and after correction.

### V. CONCLUSION

In this paper we presented a full parallel structure for RS(31,27) code encoder. This structure is suitable for high speed transmission with short code length. The test result shows that the codec works properly and can improve the error rate.

During the test we found that the built-in error detector of Reed Solomon algorithm cannot detect all the error frames. A CRC error detector is needed for further improvement.[6]

REFERENCES

[1] Ryan, W., and Lin, S. "Channel codes: classical and modern." Cambridge University Press, 2009.
[2] Hasan, M. Anwarul, and Vijay K. Bhargava. "Architecture for a low complexity rate-adaptive Reed-Solomon encoder." IEEE Transactions on Computers 44.7 (1995): 938-942.
[3] Dickson, Leonard Eugene. Linear groups: With an exposition of the Galois field theory. Courier Corporation, 2003.
[4] Reed, Irving S., and Ming-Tang Shih. "VLSI design of inverse-free Berlekamp-Massey algorithm." IEEE Proceedings E-Computers and Digital Techniques 138.5 (1991): 295-298.
[5] Sun, Q., Zhang, G., Gong, D., Deng, B., Zhou, W., You, B., Xiao, L., Wang, J, Yang, D., Liu, T., Liu, C., Guo, D., Liu, J., Hu-Guo, C., Morele, F., Valine, I., Sun, X., Ye, J., "A 3.2-Gb/s Serial Link Transmitter in 0.18 um CMOS Technology for CMOS Monolithic Active Pixel Sensors Application." IEEE Transaction on Nuclear Science, approved.
[6] Yu, C.L., Pak, E.T. and Leung, H.M., Advanced Micro Devices Inc, 1991. "ECC/CRC error detection and correction system." U.S. Patent 5,027,357.